\documentstyle[epsfig]{aipproc}
\begin{document}

\title{Prediction of Astrophysical Reaction Rates: Methods, Data Needs, and
Consequences for Nucleosynthesis Studies}

\author{Thomas Rauscher$^{1}$, Robert D.
Hoffman$^{2}$, Stanford E. Woosley$^{2,3}$, and Friedrich-Karl
Thielemann$^{1}$}

\address{$^{1}$ Dept. of Physics and Astronomy, Univ. of Basel, 4056 Basel,
Switzerland\\
$^{2}$ Lawrence Livermore National Laboratory, Livermore, CA 94550 \\
$^{3}$ Dept. of Astron. and Astroph., Univ. of California, Santa Cruz,
CA 95064}

\maketitle

\begin{abstract}
The majority of nuclear reactions in astrophysics involve unstable
nuclei which are not fully accessible by experiments yet. Therefore,
there is high demand for reliable predictions of cross sections and
reaction rates by theoretical means. The majority of reactions can
be treated in the framework of the statistical model (Hauser-Feshbach).
The global parametrizations of the nuclear properties needed for predictions
far off stability probe our understanding of the strong force and
take it to its limit.

The sensitivity of astrophysical scenarios to nuclear inputs is illustrated
in the framework of a detailed nucleosynthesis study in type II supernovae.
Abundances resulting from calculations in the same explosion model
with two different sets of reaction rates are compared. Key reactions
and required nuclear information are identified.
\end{abstract}

\section*{Introduction}
\noindent
Extensive reaction networks have to be employed in the investigation of
nuclear energy generation and nucleosynthesis processes in astrophysics. Since
stellar and explosive  burning involves a considerable number of unstable
isotopes which are currently unaccessible by experiments, the prediction of
astrophysical reaction rates by means of nuclear reaction theory is
unavoidable. There has been progress in the theoretical approaches,
especially in the modelling and prediction of specific nuclear properties
required for the determination of nuclear reaction rates. However, there is
still need for experimental studies testing the predictions and providing more
data to further improve theoretical models. After a discussion of statistical
model inputs, this will be briefly addressed in the concluding section of the
paper.

Considering the still remaining uncertainties in the prediction of
nuclear reaction rates, it is of great interest to investigate the
sensitivity of abundance yields to variations in the rates. This is
also important if one wants to disentangle stellar physics and nuclear
effects in the comparison of models which differ in both aspects.
A comparison of two sets of reaction rates by employing them in the
same stellar model of \cite{WW95} is presented in the second part of the paper.

\section*{The Statistical Model}
\noindent
In general, the cross section will be the sum of the cross sections
resulting from compound reactions via an average over overlapping
resonances (HF) and via single resonances (BW), direct reactions (DI)
and interference terms: 
\begin{equation}
\sigma (E)=\sigma ^{\mathrm{HF}}(E)+\sigma
^{\mathrm{BW}}(E)+\sigma ^{\mathrm{DC}}(E)+\sigma ^{\mathrm{int}}(E)\quad .
\end{equation}
Depending on the number of levels per energy interval in the system
projectile+target, different reaction mechanisms will dominate \cite{RTK97}.
Since different regimes of level densities are probed at the various
projectile energies, the application of a specific description depends on the
energy. In astrophysics, one is interested in energies in the range from a few
tens of MeV down to keV or even thermal energies (depending on the charge of
the projectile). 
It has been shown \cite{RTK97} that the majority of nuclear reactions
in astrophysics can be described in the framework of the statistical
model (HF) \cite{HF52}. This description assumes that the reaction
proceeds via a compound nucleus which finally decays into the reaction
products. With a sufficiently high level density, average cross sections
\begin{equation}
\sigma ^{\textrm{HF}}=\sigma
_{\textrm{form}}b_{\textrm{dec}}=\sigma _{\textrm{form}}{\Gamma
_{\textrm{final}}\over \Gamma _{\textrm{tot}}}
\end{equation}
can be calculated which can be
factorized into a cross section $\sigma _{\textrm{form}}$ for the formation of
the compound nucleus and a branching ratio $b_{\textrm{dec}}$, describing the
probability of the decay into the channel of interest compared with the total
decay probability into all possible exit channels. The partial widths $\Gamma$ 
as well as $\sigma _{\textrm{form}}$ are related to (averaged) transmission
coefficients, which comprise the central quantities in any HF calculation.

Many nuclear properties enter the computation of the transmission coefficients:
mass differences (separation energies), optical potentials, GDR widths,
level densities. The transmission coefficients can be modified due
to pre-equi\-li\-bri\-um effects which are included in width fluctuation
corrections \cite{T74} (see also \cite{RTK97} and references therein)
and by isospin effects. It is in the description of the nuclear properties
where the various HF models differ. A choice of what is thought of
being the currently best parametrizations is incorporated in the new
HF code NON-SMOKER \cite{rt98}, which is based on the well-known code
SMOKER \cite{thi87}.

\section*{A Reaction Rate Library}
\noindent
Utilizing the NON-SMOKER code, cross sections and reaction rates for
reactions with nucleons, $\alpha$  particles or $\gamma$  rays in entrance
and exit channels, respectively, were calculated for all targets between
proton and neutron drip line in the range $9<Z<84$. Tabulated cross
sections and rates can be found at
{\em http://quasar.physik.unibas.ch/\~{ }tommy/reaclib.html}. Analytic fits to
these rates, along with further information, can be obtained as an electronic
file from the authors or on-line from {\em Atomic Data and Nuclear Data
Tables}. A selection from these fits is published in \cite{RT00}.

In all applications, these rates should be supplemented or replaced
with experimental rates as they become available. Such a combination
of theoretical and experimental rates is provided, e.g., in the REACLIB
compilation. Currently, a new version is being compiled, in which
the theoretical rates presented here will be included. Latest information
on REACLIB can be found on the WWW at {\em http://ie.lbl.gov/astro.html}.
Further details on the NON-SMOKER code are presented at
{\em http://quasar.physik.unibas.ch/\~{ }tommy/reaclib.html}.

\section*{Reaction Rate Sensitivity of Nucleosynthesis in Type II Supernovae}
\noindent
When comparing results from different supernova models one faces the
difficulty caused by the fact that it is hard to differentiate between
influences of differing reaction rate sets and different stellar physics.
We tried to segregate the abundance differences between the two models
of \cite{WW95} (WW95) and \cite{TNH96} (TNH) existing because of the
dichotomy of stellar models from those reflecting purely the choice
of nuclear physics. For that purpose, hybrid calculations were performed,
using the same stellar evolution code as in \cite{WW95} but with rates
from both models. In addition to helping to understand why calculations
of the two groups differ, the use of independent rate sets in identical
stellar models helps determine the nuclear physics portion of the
error bar one should assign to nucleosynthesis studies of this sort.
The cause of the differences in the theoretical rates
was further investigated, pointing to possibilities for future improvements
of rate predictions. In the following, the findings are briefly summarized.
A very detailed account of the work can be found in \cite{HWWRT99}.

\subsection*{The Rate Sets}
\noindent
The reaction rates utilized in WW95 were those of \cite{WFH78} (WFHZ),
TNH used \cite{thi87} (TAT). As examples, Figs.\ 1 and 2 show a comparison
of the two sets to each other and to experimental values for 30 keV
neutron capture and proton capture at $T_{9}=3$.
\begin{figure}[t]
\centerline{\epsfig{file=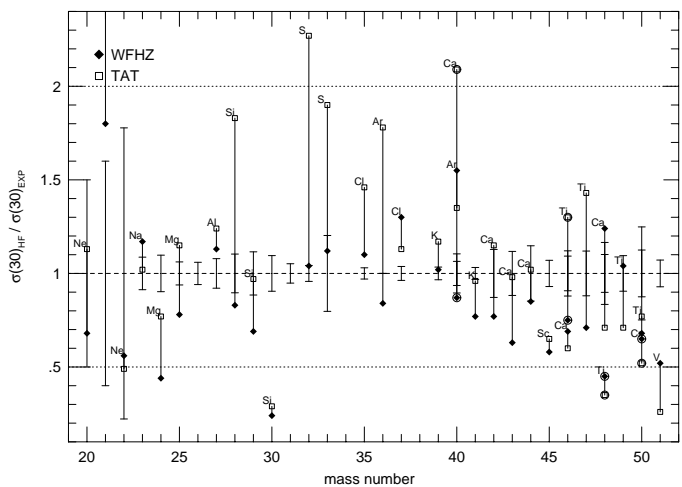,height=10.5cm}} 
%\vspace{10pt} 
\caption{30 keV neutron capture cross sections.} 
\end{figure}
\begin{figure}[t]
\centerline{\epsfig{file=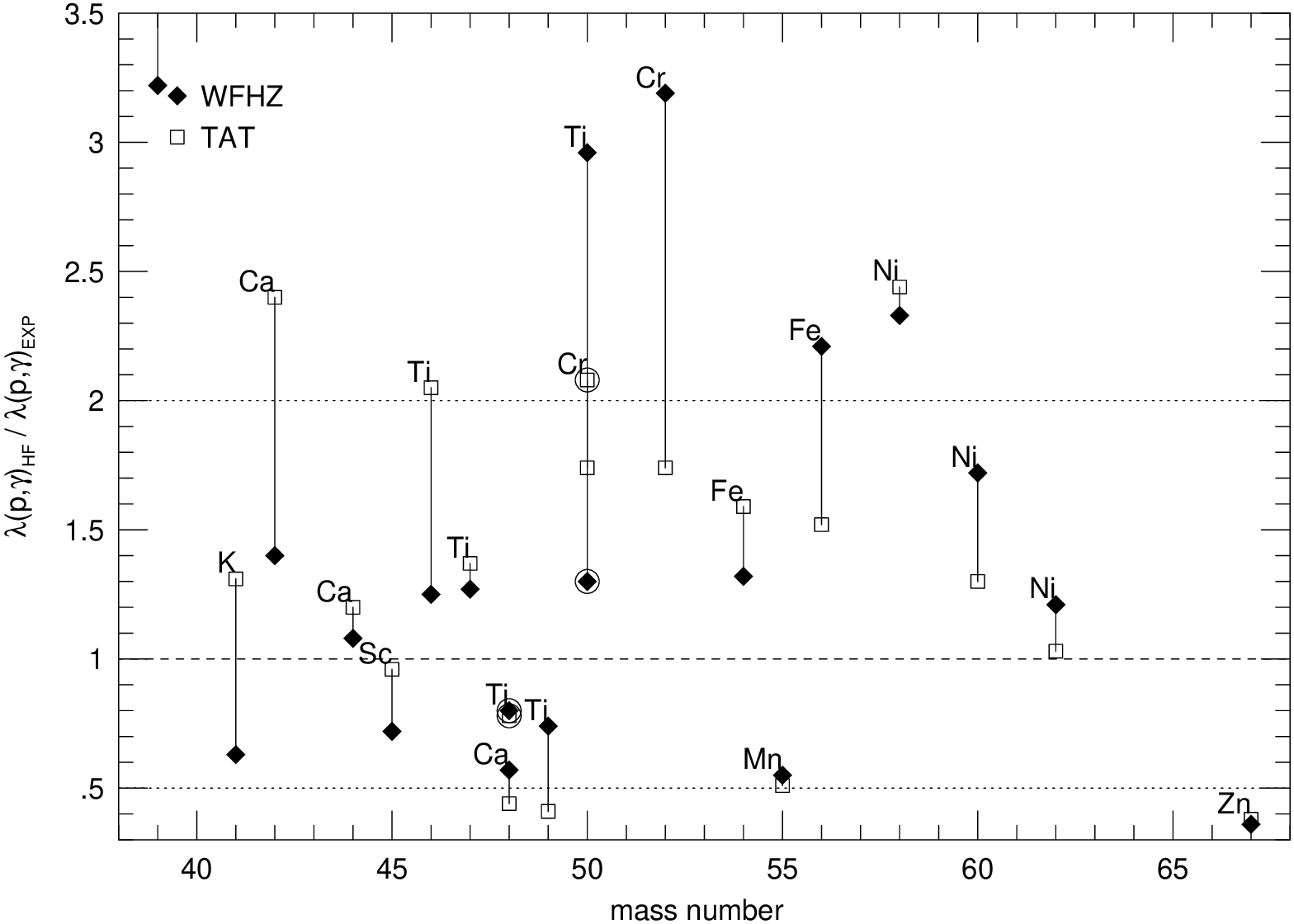,height=10.5cm}} 
%\vspace{10pt} 
\caption{Proton capture cross sections at $T_9=3$.} 
\end{figure}

Typical differences at astrophysically interesting temperatures are
less than a factor of two. There are individual cases, however, where
the difference exceeds a factor of 10. Some of the larger differences
occur for reactions where scarce experimental information is available
and different assumptions were made regarding the photon transmission
function, for example, ($\alpha$,$\gamma$) reactions on $Z = N$ nuclei.
Different assumptions were also made about the particle transmission
functions, nuclear partition functions, and level densities. More
modern and complete data used in the TAT rates makes them superior
in cases where the partition function is important. WFHZ used an equivalent
square well with empirical reflection factors; TAT used a more detailed
optical model. Given the quite different values for, e.g., the neutron
and proton transmission function, it is perhaps surprising that the
rates differ so little. This is because the relevant temperatures
for explosive burning are high. For incident particles in the Gamow
window, the deviations in the particle transmission functions are
typically smaller than a factor of two. In addition, higher partial
waves contribute. A comparison of rates at a lower temperature would
have revealed larger discrepancies.

Compared to experiment, both sets of theoretical rates give similar
agreement, typically to a factor of two. The standard deviations between
the two theoretical sets and cross section data are almost identical.
In summary, the two rate sets have comparable merit when compared
to experiment.
All the authors of this paper agree that the new rate set, the ``NON-SMOKER''
set, will be preferable to both TAT and WFHZ and will be adopted by
both groups (WW and TNH) for future work.

\subsection*{Results and Conclusions}
\noindent
The comparison of the yields obtained with the two reaction rate sets
in the WW95 model is shown in Figs.\ 3 and 4 for a 15 and a 25 $M_{\odot }$
supernova, respectively.
\begin{figure}[t]
\centerline{\epsfig{file=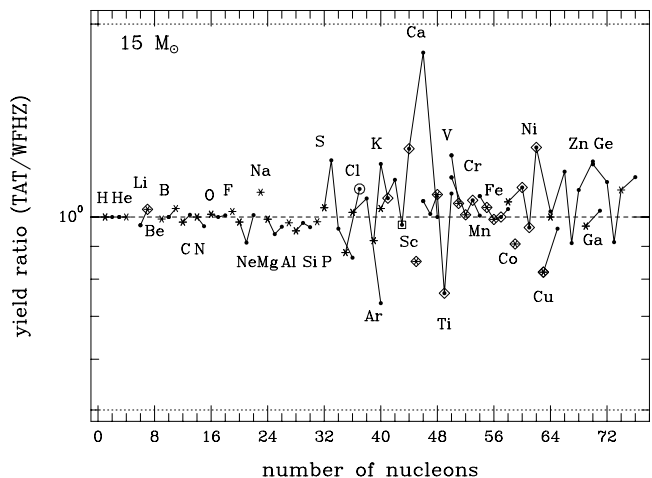,height=11cm}} 
%\vspace{10pt} 
\caption{Yield ratios for a 15 $M_{\odot }$ supernova.} 
\end{figure} 
  
When the two current rate sets
are included in otherwise identical stellar models we find that the
nucleosynthesis, with some interesting exceptions, is not greatly
changed. For example, only about a dozen (out of 70) stable isotopes
in the mass range 12 to 70 have nucleosynthesis that differ by over
20\% in two supernovae of 15 $M_{\odot }$ that use the same rate for
$^{12}$C($\alpha ,\gamma$)$^{16}$O. It can, however, be noticed that most
of these isotopes - with one exception $^{44}$Ti - are products of hydrostatic
burning where individual reaction flows are governed by the cross
sections involved. Nevertheless, none differ by more than a factor
of 1.7. Given the significantly larger differences that exist in individual
reaction rates, one may wonder at the robust nature of the final nucleosynthesis.
We see three major causes.
\begin{figure}[t]
\centerline{\epsfig{file=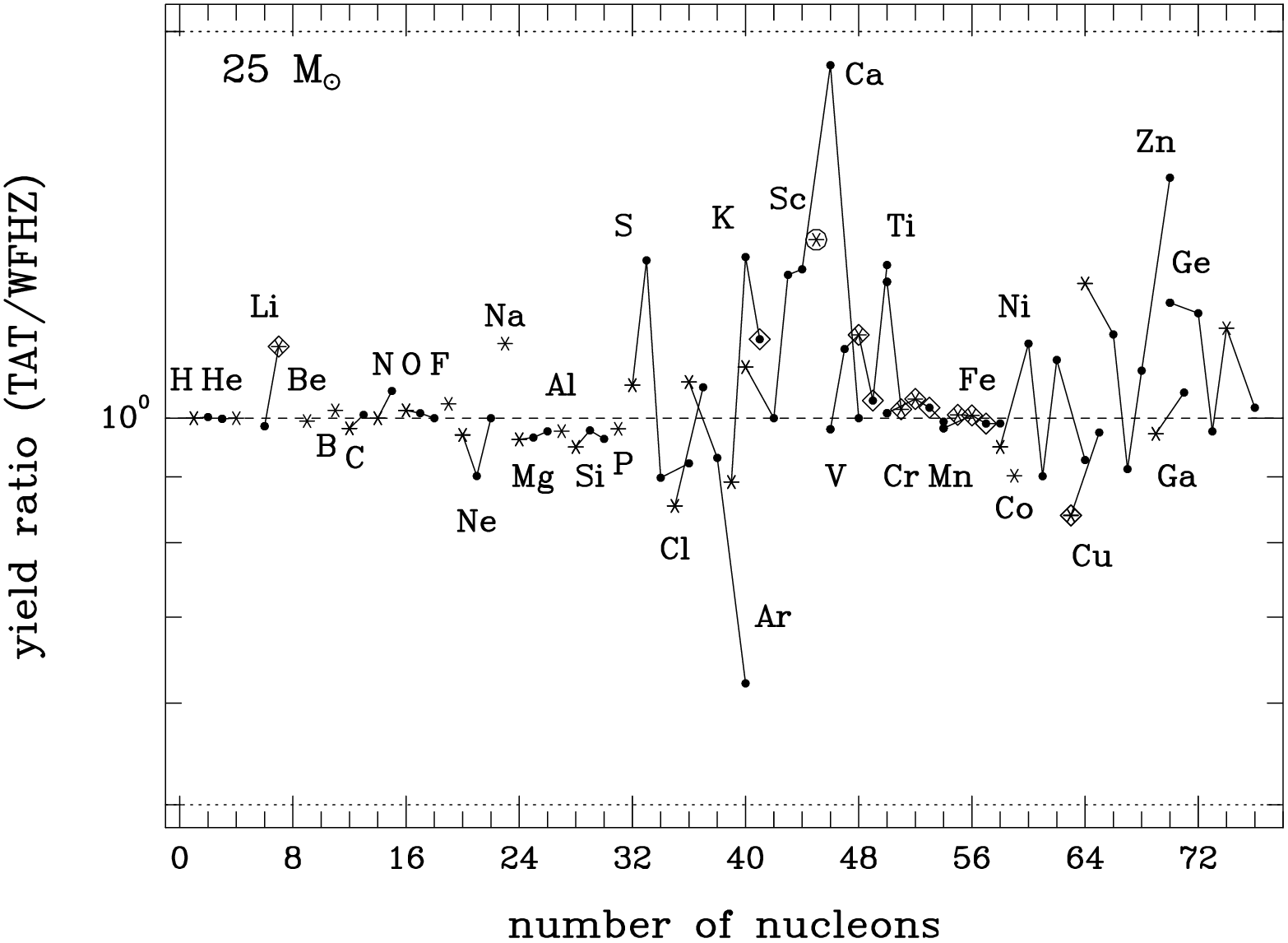,height=11cm}} 
%\vspace{10pt} 
\caption{Yield ratios for a 25 $M_{\odot }$ supernova.} 
\end{figure} 

First, as the star burns and becomes hotter, the nuclear flow follows
the valley of $\beta$  stability making heavier nuclei as it goes. In
doing so, it follows the path of least resistance -- those reactions
having the largest cross section for a nucleon or $\alpha$  particle
reacting with a given nucleus. These large cross sections are reasonably
well replicated by any calculation, normalized to experiment, that
treats the Coulomb barrier and photon transmission function approximately
correctly. Large differences may exist in rate factors for reactions
that are in competition, especially a small channel in the presence
of a large one, but these small channels are frequently negligible,
at least for the major abundances while they can cause larger differences
when one is interested in the abundances of trace isotopes, e.g.\ in isotopic
anomalies of meteoritic inclusions.

If one is, however, interested in accurate abundance predictions resulting
from these smaller flows in hydrostatic burning stages, these can in
most cases only be obtained by improving the reliability of the cross
sections (and reaction rates) that determine these weak flows on light
and intermediate mass nuclei. As new experimental information becomes
available, a continuous improvement is therefore highly warranted.

Second, beyond oxygen burning (nuclei heavier than
calcium), nucleosynthesis increasingly occurs in a state of full or
partial nuclear statistical equilibrium. There the abundances are
given by binding energies and partition functions. As long as the
``freeze-out'' is sufficiently rapid, individual rates are not so important.

Third, the reaction rates varied here were only those theoretical values
from Hauser-Feshbach calculations for intermediate mass nuclei, i.e.,
nuclei heavier than magnesium. The really critical reaction rates
are, for the most part, those below magnesium. These reactions, like
e.g., $^{12}$C($\alpha ,\gamma$)$^{16}$O, govern the energy generation,
major nucleosynthesis, and neutron exposure in the star. The rest
are perturbations on these dominant flows.

This is not to say, however, that the nuclear and stellar details of
heavy element synthesis are now well understood. Differences in the
stellar model may account not just for 20\% variation, but orders of
magnitude. That is, uncertainty in stellar physics -- especially the
treatment of convection and how it is coupled (or not coupled) to
the nuclear network -- accounts for most of the differences in current
nucleosynthesis calculations -- provided such calculations use the
same nuclear reaction rates below magnesium.

Even in a perfect stellar model though, there will still be interesting
nuclear physics issues. Stellar nucleosynthesis is becoming a mature
field rich with diverse and highly detailed observational data. The
``factor of two'' accuracy that was adequate in the past may not do
justice to the observations of the future. There are many individual
cases where the nuclear physics uncertainty is still unacceptably
large. We point out just a few examples for which experimental information
would be of interest.

\section*{Nuclear data needs}
\noindent
The suppression of radiative capture reactions into self-conjugate
(isospin zero) nuclei is very uncertain. Past Hauser-Feshbach calculations
have adopted empirical factors for this suppression. The new NON-SMOKER
rates include a significantly improved treatment \cite{RGW99}. $\alpha
$-capture reactions, like $^{24}$Mg($\alpha ,\gamma $)$^{28}$Si,
$^{28}$Si($\alpha ,\gamma $)$^{32}$S, \ldots{}, $^{44}$Ti($\alpha ,\gamma
$)$^{48}$Cr, are very important to nucleosynthesis in oxygen and silicon
burning. The reaction $^{40}$Ca($\alpha ,\gamma $)$^{44}$Ti also directly
affects the synthesis of $^{44}$Ti. Modern accurate determinations of most of
the reaction rates are missing (as well as (p,$\gamma $) reactions into the
same nuclei). Measurements here would be most welcome.

The Hauser-Feshbach rates are also only as good as the local experimental
rates to which the necessary parameters of the calculation are calibrated.
In that regard we would point out the near absence of charged particle
reaction rate data for $A>70$, even for stable nuclei. Charged particle
reactions are important, especially on unstable nuclei, at significantly
higher atomic weights in the $r$ process and in the $p$ process.

Extended systematics of other nuclear properties, such as level densities
(especially around magic nucleon numbers), low energy behavior of the GDR, and
optical potentials, would be highly appreciated for {\em stable} as well as
unstable nuclei.

\subsection*{Acknowledgements}
\noindent
This work has been supported by the U.S. NSF (AST 96-17494, AST
96-17161, AST 97-31569, PHY-74-07194) and the Swiss NSF (20-47252.96,
2000-53798.98). T. R. is a PROFIL fellow of the Swiss NSF.

\newcommand{\noopsort}[1]{} \newcommand{\printfirst}[2]{#1}
  \newcommand{\singleletter}[1]{#1} \newcommand{\swithchargs}[2]{#2#1}


\begin{thebibliography}{10}

\bibitem{WW95}
S.~E. Woosley and T.~A. Weaver, Ap. J. Suppl. {\bf 101},  181  (1995).

\bibitem{RTK97}
T. Rauscher, F.-K. Thielemann, and K.-L. Kratz, Phys.\ Rev.\ C {\bf 56},  1613
  (1997).

\bibitem{HF52}
W. Hauser and H. Feshbach, Phys. Rev. A {\bf 87},  366  (1952).

\bibitem{T74}
J. Tepel, H. Hoffmann, and H. Weidenm{\"u}ller, Phys.\ Lett. {\bf 49B},  1
  (1974).

\bibitem{rt98}
T. Rauscher and F.-K. Thielemann,  in {\em Stellar Evolution, Stellar
  Explosions and Galactic Chemical Evolution}, edited by A. Mezzacappa (IOP,
  Bristol, 1998), p.\ 519.

\bibitem{thi87}
F.-K. Thielemann, M. Arnould, and J.~W. Truran,  in {\em {A}dvances in
  {N}uclear {A}strophysics}, edited by E. Vangioni-Flam {\it et~al.} (Editions
  Fronti\`ere, Gif sur Yvette, 1987), p.\ 525.

\bibitem{RT00}
T. Rauscher and F.-K. Thielemann, At.\ Data Nucl.\ Data Tabl., submitted
  (1999).

\bibitem{TNH96}
F.-K. Thielemann, K. Nomoto, and M. Hashimoto, Ap.\ J. {\bf 460},  408  (1996).

\bibitem{HWWRT99}
R.~D. Hoffman {\it et~al.}, Ap.\ J. {\bf 521},  735  (1999).

\bibitem{WFH78}
S. Woosley, W. Fowler, J. Holmes, and B. Zimmerman, At.\ Data Nucl.\ Data
  Tabl. {\bf 22},  371  (1978).

\bibitem{RGW99}
T. Rauscher, J. G{\"o}rres, and M.~C. Wiescher, Phys.\ Rev.\ C  , to be
  submitted  (1999).

\end{thebibliography}
\end{document}